\documentclass[twocolumn]{aastex63}
\let\tablenum\relax
\usepackage{siunitx}
\usepackage{float,graphicx,amsmath,multirow}
\DeclareSIUnit\molecule{molecule}
\usepackage[version=3]{mhchem}
\usepackage{booktabs}
\usepackage{enumitem}
\usepackage{blindtext}

\newcommand\newtag[2]{#1\def\@currentlabel{#1}\label{#2}}

\submitjournal{ApJ}
\turnoffeditone
\turnoffedittwo
\shorttitle{Discovery of cyano-butadiene with GOTHAM in TMC-1}
\shortauthors{Cooke et al.}
\graphicspath{{./}}
\begin{document}
%TC:ignore

\title{Detection of Interstellar \textit{E}-1-cyano-1,3-butadiene in GOTHAM Observations of TMC-1}

%and Implications for the Formation of Aromatic Molecules

\author[0000-0002-0850-7426]{Ilsa R. Cooke}
\affiliation{Department of Chemistry, University of British Columbia, 2036 Main Mall, Vancouver, BC V6T 1Z1, Canada}
%\email{icooke@chem.ubc.ca}

\author[0000-0003-2760-2119]{Ci Xue}
\affiliation{Department of Chemistry, Massachusetts Institute of Technology, Cambridge, MA 02139, USA}

\author[0000-0003-0304-9814]{P. Bryan Changala}
\affiliation{Center for Astrophysics $\mid$ Harvard~\&~Smithsonian, Cambridge, MA 02138, USA}
\author{Hannah Toru Shay}
\affiliation{Department of Chemistry, Massachusetts Institute of Technology, Cambridge, MA 02139, USA}

\author[0000-0002-4593-518X]{Alex N. Byrne}
\affiliation{Department of Chemistry, Massachusetts Institute of Technology, Cambridge, MA 02139, USA}

\author{Qi Yu Tang}
\affiliation{Department of Chemistry, University of British Columbia, 2036 Main Mall, Vancouver, BC V6T 1Z1, Canada}

\author{Zachary T. P. Fried}
\affiliation{Department of Chemistry, Massachusetts Institute of Technology, Cambridge, MA 02139, USA}

\author[0000-0002-1903-9242]{Kin Long Kelvin Lee}
\affiliation{Accelerated Computing Systems and Graphics Group, Intel Corporation, 2111 25th NE Ave, Hillsboro, OR 97124, USA}
\affiliation{Department of Chemistry, Massachusetts Institute of Technology, Cambridge, MA 02139, USA}

\author{Ryan A. Loomis}
\affiliation{National Radio Astronomy Observatory, Charlottesville, VA 22903, USA}

\author[0000-0001-6705-2022]{Thanja Lamberts}
\affiliation{Leiden Institute of Chemistry, Gorlaeus Laboratories, Leiden University, P.O. Box 9502, 2300 RA Leiden, The Netherlands}
\affiliation{Leiden Observatory, Leiden University, P.O. Box 9513, 2300 RA Leiden,
The Netherlands}

\author[0000-0001-9479-9287]{Anthony Remijan}
\affiliation{National Radio Astronomy Observatory, Charlottesville, VA 22903, USA}

\author[0000-0003-0799-0927]{Andrew M. Burkhardt}
\affiliation{Department of Physics, Wellesley College, Wellesley, MA 02481, USA}

\author{Eric Herbst}
\affiliation{Department of Chemistry, University of Virginia, Charlottesville, VA 22904, USA}
\affiliation{Department of Astronomy, University of Virginia, Charlottesville, VA 22904, USA}

\author[0000-0001-9142-0008]{Michael C. McCarthy}
\affiliation{Center for Astrophysics $|$ Harvard \& Smithsonian, Cambridge, MA 02138, USA}

\author[0000-0003-1254-4817]{Brett A. McGuire}
\affiliation{Department of Chemistry, Massachusetts Institute of Technology, Cambridge, MA 02139, USA}
\affiliation{National Radio Astronomy Observatory, Charlottesville, VA 22903}

\correspondingauthor{Ilsa R. Cooke, Brett A. McGuire}
\email{icooke@chem.ubc.ca, brettmc@mit.edu}

\begin{abstract}

We report the detection of the lowest energy conformer of \textit{E}-1-cyano-1,3-butadiene ($E$-1-\ce{C4H5CN}), a linear isomer of pyridine, using the fourth data reduction of the GOTHAM deep spectral survey toward TMC-1 with the 100 m Green Bank Telescope. We performed velocity stacking and matched filter analyses using Markov chain Monte Carlo simulations and find evidence for the presence of this
molecule at the 5.1$\sigma$ level. We derive a total column density of $3.8^{+1.0}_{-0.9}\times 10^{10}$ cm$^{-2}$, which is predominantly found toward two of the four velocity components we observe toward TMC-1. We use this molecule as a proxy for constraining the gas-phase abundance of the apolar hydrocarbon 1,3-butadiene. Based on the three-phase astrochemical modeling code \texttt{NAUTILUS} and an expanded chemical network, our model underestimates the abundance of cyano-1,3-butadiene by a factor of 19, with a peak column density of $2.34 \times 10^{10}\ \mathrm{cm}^{-2}$ for 1,3-butadiene. %Using the \ce{1-C4H5CN}:\ce{CH2CHCHCH2} ratio derived in our model and the observed column density of \ce{1-C4H5CN}, we further constrain the total column density of 1,3-butadiene in TMC-1 to $3.94 \times 10^{11}\ \mathrm{cm}^{-2}$.
Compared to the modeling results obtained in previous GOTHAM analyses, the abundance of 1,3-butadiene is increased by about two orders of magnitude. Despite this increase, the modeled abundances of aromatic species do not appear to change and remain underestimated by 1--4 orders of magnitude. Meanwhile, the abundances of the five-membered ring molecules increase proportionally with 1,3-butadiene by two orders of magnitudes.  We discuss implications for bottom-up formation routes to aromatic and polycyclic aromatic molecules. 

\end{abstract}

%TC:endignore

\section{Introduction}
\label{sec:intro}

The dark prestellar cloud TMC-1, in the Taurus Molecular Cloud complex, has been extensively observed revealing a rich molecular inventory, including carbon-chain neutral species, positive and negative ions, and nitrogen-bearing species \citep{Gratier:2016fj}. More recently, the discovery of benzonitrile (c-C$_6$H$_5$CN) by \citet{McGuire:2018it} added the first aromatic ring to this inventory. Following this detection, a number of additional aromatic and unsaturated cyclic molecules have been detected using our GOTHAM (GBT Observations of TMC-1: Hunting for Aromatic Molecules) line survey \citep{McCarthy:2021aa,McGuire2021,Burkhardt2021,Lee:2021ud} and the QUIJOTE (Q-band Ultrasensitive Inspection Journey to the Obscure TMC-1 Environment) \edit3{survey of Cernicharo and colleagues\citep{Cernicharo2021,Cernicharo2021a,Cernicharo2021b, Cernicharo2022}}.

In order to understand the formation of aromatic molecules and other cyclic species, gas-grain chemical models (e.g. \texttt{NAUTILUS}; \citealt{Ruaud:2016}) combined with large reaction networks (e.g. \citealt{Wakelam:2015dr}) have been used. These models, however, cannot currently reproduce the observed abundance of these aromatic molecules, which may be in part due to the lack of observational constraints on potential precursors \citep{Burkhardt:2021aa}. Small, unsaturated hydrocarbons are suggested to be involved in the bottom-up formation of aromatic molecules. However, many unsaturated and partially saturated hydrocarbons have not yet been detected in TMC-1 and thus are large unknowns for these models. As such, in order to determine the formation pathways of even the simplest aromatics, robust abundance measurements of potential precursor species must be obtained. 

In laboratory experiments under single-collision conditions, \citet{Jones:2011yc} showed that benzene (\ce{C6H6}) can readily form via the neutral-neutral reaction between the ethynyl radical (C$_2$H) and 1,3-butadiene (\ce{CH2CHCHCH2}):
\begin{equation}
    \ce{CH2CHCHCH2 + C2H -> C6H6 + H}.
\end{equation}
Electronic structure calculations showed the reaction is barrierless and exoergic, forming \ce{C6H6} through a complex-forming reaction mechanism. The thermodynamically less stable hexa-1,3-dien-5-yne isomer (\ce{HCCCHCHCHCH2}) of benzene was found to be the dominant reaction product under single collision conditions, whereas the branching ratio to \ce{C6H6} was 30\%$\pm$10\%.

\citet{Lockyear:2015dn} studied the products of this reaction under thermal conditions in a flow reactor using synchrotron photoionization mass spectrometry. The photoionization spectra indicated, in contrast to the observations of \citet{Jones:2011yc}, that the fulvene (\ce{(CHCH)2CCH2}) isomer of benzene is the major reaction product, with a branching fraction of $\sim$60\%. They did not detect \ce{C6H6} as a product at all and placed an upper limit on the branching fraction for the sum of the \ce{C6H6} and \ce{HCCCHCHCHCH2} isomers of 45\%. \citet{Lee:2019bc} likewise observed evidence for \ce{(CHCH)2CCH2} formation in a microwave discharge containing \ce{HC3N} (used as a \ce{C2H} precursor) and \ce{CH2CHCHCH2}; \ce{C6H6} could not be constrained in their work since rotational spectroscopy was used as the detection method. 

\ce{CH2CHCHCH2} is therefore a critical yet unconstrained precursor to aromatic and cyclic molecules in TMC-1. However, since the lowest energy conformer of \ce{CH2CHCHCH2} does not possess a dipole moment, it is invisible to radio astronomy. Searching instead for an analog of that molecule which has been ``tagged"  with a polar functional group, such as the nitrile (or cyano) unit, -\ce{C#N}, yields a spectroscopically bright surrogate. Indeed, the first detections of a benzene-ring species with radio astronomy, and individual interstellar polycyclic aromatic hydrocarbon (PAH) molecules in general, were those of the nitrile derivatives of \ce{C6H6} and naphthalene (\ce{C10H8}), namely benzonitrile \citep[\ce{C6H5CN};][]{McGuire:2018it} and cyanonapthalene \citep[1- and 2-\ce{C10H7CN};][]{McGuire2021}, respectively. Laboratory and theoretical studies indicate that nitrile functionalization of unsaturated hydrocarbons occurs facilely via CN-addition and H-elimination across a double bond \citep{Sims1993,Balucani:2000nw}. This makes the observation of CN-substituted surrogates a potentially broadly applicable tool for quantifying otherwise radio-dark symmetric hydrocarbons. Recent observations and astrochemical models, comparing the ratio of pure aromatic hydrocarbons to their CN-functionalized counterparts support this proxy method \citep{Sita_2022}.

The reaction of CN with \ce{CH2CHCHCH2} has been studied extensively both in the laboratory and computationally \citep{2014JPCA..118.7715S,Morales:2011gl}. \citet{Morales:2011gl} used crossed molecular beams to investigate the reaction products under single-collision conditions and uniform supersonic flows to measure the reaction kinetics as a function of temperature. Their measurements demonstrated that at low temperatures the overall reaction is fast, close to the gas-kinetic limit and that the 1-cyano-1,3-butadiene (\ce{1-C4H5CN}) isomer is the dominant reaction product. They also suggested that possible minor fractions of the aromatic pyridine isomer (\ce{c-C5H5N}) could form in this reaction, a heterocycle which has thus far eluded detection in the ISM \citep{Barnum2022}. We were therefore motivated to search for \ce{1-C4H5CN} in TMC-1 as a chemical proxy for \ce{CH2CHCHCH2}.   Recently, \citet{Zdanovskaia2021} reported millimeter-wave rotational spectra of three isomers of cyanobutadiene:
the \textit{s-trans} conformers of \textit{E}-1-cyano-1,3-butadiene, and \textit{Z-}1-cyano-1,3-butadiene, and the \textit{syn} and \textit{anti} conformers of 4-cyano-1,2-butadiene, extending the centimeter-wave measurements of the \textit{E-} and \textit{Z-} steroisomers reported by \citet{McCarthy2020}. These new laboratory measurements, facilitated by the synthesis of these molecules \citep{Kougias2020}, provide the foundation for our astronomical search.

\begin{figure}
\centering
\includegraphics[width=2.5in]{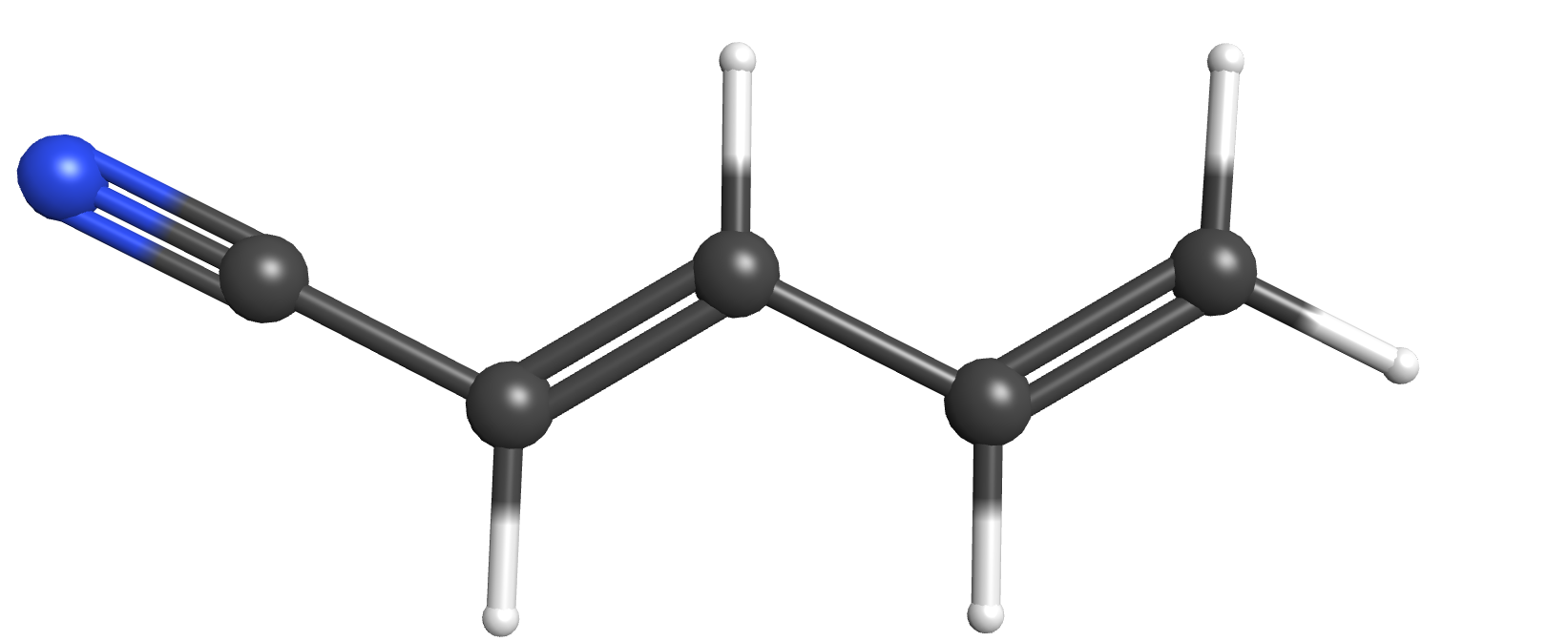}
\caption{\label{fig:structure} The chemical structure of \textit{s-trans-E}-1-cyano-1,3-butadiene.} 
\end{figure}

We, therefore, perform a dedicated astronomical search for cyano-butadiene isomers with the GOTHAM spectral survey. The observations are described in Section~\ref{sec:obs}. Section~\ref{sec:stacking} presents the observational analyses and the detection of \textit{s-trans-E}-1-cyano-1,3-butadiene, which we hereafter refer to as \ce{$E$-1-C4H5CN} for simplicity. The chemical structure of the \textit{s-trans} conformer is shown in Figure \ref{fig:structure}. The observed results are used to constrain the new chemical network developed for \ce{1-C4H5CN} and its isomer \ce{c-C5H5N} in Section~\ref{sec:model}. In Section~\ref{sec:dis}, we discuss the implications of constraining the abundance of \ce{1-C4H5CN} and the future searches for \ce{c-C5H5N}. Finally, we summarize our results in Section~\ref{sec:con}.

\section{Observations}
\label{sec:obs}

We performed a search toward TMC-1 using the fourth data reduction of the GOTHAM collaboration survey. GOTHAM is a large project performed with the 100 m  Robert C. Byrd Green Bank Telescope (GBT). The GOTHAM program is a dedicated spectral line observing program of TMC-1 covering almost 30 GHz of bandwidth at high sensitivity and spectral resolution. Details of the source, observations, and data reduction methods can be found in \citet{McGuire:2020bb} and \citet{McGuire2021}. Briefly, the spectra in these data cover the entirety of the X-, K-, and Ka-receiver bands with nearly continuous coverage from 7.9 to 11.6 GHz, 12.7 to 15.6 GHz, and 18.0 to 36.4 GHz (24.9 GHz of total bandwidth). Observations were performed with the project codes GBT17A\nobreakdash-164,  GBT17A\nobreakdash-434, GBT18A\nobreakdash-333, GBT18B\nobreakdash-007, GBT19B\nobreakdash-047, AGBT20A-516, AGBT21A-414, and AGBT21B-210.

The first, second, and third data reduction of GOTHAM (hereafter referred to as DR1, DR2 and DR3) comprise observations obtained between February 2018 -- May 2019 (DR1), June 2020 (DR2), and April 2021 (DR3) \citep{McGuire:2020bb,McGuire2021}. The GOTHAM observations used here are the fourth data reduction (DR4), which comprises observations made through May 2022. DR4 extends the frequency coverage to 7.906 -- 36.411 GHz (24.9 GHz total observed bandwidth) with a few gaps and improved the sensitivity in some frequency coverage already covered by DR2. A full description of DR4 can be found in \citet{Sita_2022}. 

The pointing was centered on the TMC\nobreakdash-1 cyanopolyyne peak (CP) at (J2000) $\alpha$~=~04$^h$41$^m$42.50$^s$ $\delta$~=~+25$^{\circ}$41$^{\prime}$26.8$^{\prime\prime}$. The spectra were obtained through position-switching to an emission-free position 1$^{\circ}$ away. Pointing and focusing were refined every 1--2 hours, primarily on the calibrator J0530+1331. Flux calibration was performed with an internal noise diode and Karl G. Jansky Very Large Array (VLA) observations of the same calibrator used for pointing, resulting in a flux uncertainty of ${\sim}20$\% \citep{McGuire:2020bb,Sita_2022}. All data were taken with a uniform frequency resolution of 1.4\,kHz (0.05--0.01\,km/s in velocity). An RMS noise of ${\sim}2$--20\,mK was achieved across most of the observed frequency range with the RMS gradually increasing toward higher frequencies due to less total integration time at those settings.  

\section{Analysis and Results}

\subsection{Observational Analyses and Results}
\label{sec:stacking}

As of GOTHAM DR4, no individual transitions of the cyano-butadiene isomers were bright enough for identification. In the absence of strong individual lines, we used the spectral stacking and matched filtering procedure detailed in \citet{Loomis:2021aa} to determine the statistical evidence for the presence of this molecule. In summary, a small spectral window is extracted for each of the top 5\% strongest predicted transitions, provided that there is no interloping emission ($>$5$\sigma$) present in the spectrum. Fiducial stacks for \textit{Z-}1-cyano-1,3-butadiene did not reveal any signal of significance, whereas signal was observed for the \textit{E-}isomer. In this case, a total of 434 rotationally-resolved hyperfine transitions of \ce{1-C4H5CN} met these criteria with no interloping transitions detected. The windows were subsequently combined in velocity space, each weighted by the observational RMS and the predicted flux.

We additionally carry out a forward modeling procedure using \texttt{molsim} \citep{lee_molsim_2020} that simulates the molecular emission with a set of model parameters following the conventions of \citet{Turner:1991:617} for a single excitation temperature and accounting for the effect of optical depth.  The parameters include the source size (SS)---used for estimating beam dilution effects---radial velocity ($v_{lsr}$), column density ($N_T$), excitation temperature ($T_{ex}$), and the line width ($\Delta $V). \edit1{More details of our modelling approach can be found in \citet{Loomis:2021aa}}

Prior high-resolution observations of TMC-1 from both from GOTHAM \citep{Xue:2020:L9} and others \citep{Dobashi:2018:82,Dobashi:2019:88} have found that most emission seen at cm-wavelengths can be separated into contributions from four distinct velocity components within the larger structure, at approximately 5.4, 5.6, 5.8, and 6.0\,km\,s$^{-1}$ \citep{Loomis:2021aa}. In our model, these are assigned to independent sets of SS, $v_{lsr}$, and $N_{T}$, while a uniform $T_{ex}$ and $\Delta $V are adopted, resulting in a total of 14 modeling parameters. To properly account for uncertainty and covariance between model parameters, we use Affine-invariant Markov Chain Monte Carlo (MCMC) sampling as implemented in \texttt{emcee}, estimating the likelihood for each given set of model parameters. As with previous work on aromatic and cyclic species \citep{McCarthy:2021aa, Lee:2021ud,McGuire2021}, our choice of prior was on the basis of chemical similarity; the posterior for HC$_9$N was used as the prior distribution for the \ce{1-C4H5CN} MCMC modeling.

\begin{figure*}[bt]
    \centering
    \includegraphics[width=0.45\textwidth]{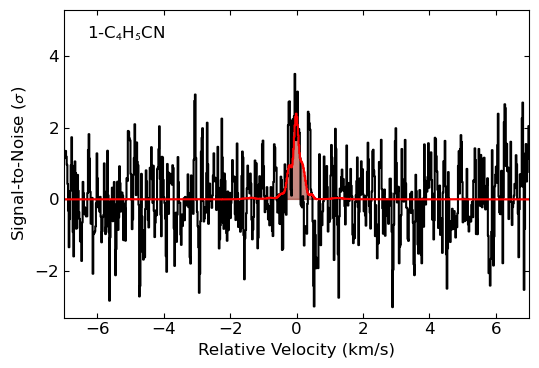}
    \includegraphics[width=0.45\textwidth]{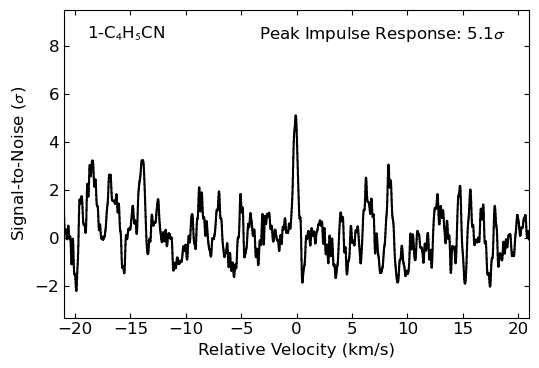}
    \caption{Velocity stacked and matched filter spectra of \ce{$E$-1-C4H5CN}. The intensity scales are the signal-to-noise ratios (SNR) of the response functions when centered at a given velocity. The ``zero'' velocity corresponds to the channel with the highest intensity to account for blended spectroscopic transitions and variations in velocity component source sizes. (\emph{Left}) The stacked spectra from the GOTHAM DR4 data are displayed in black, overlaid with the expected line profile in red from our MCMC fit to the data.  The signal-to-noise ratio is on a per-channel basis. (\emph{Right}) Matched filter response obtained from cross-correlating the simulated and observed velocity stacks in the left panel; value annotated corresponds to the peak impulse response of the matched filter.}
    \label{fig:stack}
\end{figure*}

Following convergence of the sampling, the resulting posterior was analyzed using the \texttt{arviz} suite of routines. To assess the robustness of the detection, the simulated spectrum from the posterior mean is velocity stacked and cross-correlated with the observational velocity stack in a matched filter analysis: the peak impulse response ($\sigma$) corresponds to the statistical significance. As with previous GOTHAM analyses, we adopt a $5\sigma$ threshold to classify a firm detection \citep{Loomis:2021aa}.

\begin{table}[bt]
    \centering
        \caption{Summary statistics of the marginalized \ce{$E$-1-C4H5CN} posterior. The quoted uncertainties correspond to the 16th and 84th percentiles (1$\sigma$ for a Gaussian distribution). The total column density is derived from combining the column densities of each component in quadrature. }
    \begin{tabular}{ c c c c c}
    \toprule
    	$v_{lsr}$	&	Size	&	$N_T$	&	$T_{ex}$	&	$\Delta V$		\\
    	(km\,s$^{-1}$)&	($^{\prime\prime}$)	&	(10$^{10}$cm$^{-2}$) & (K)	&  (km\,s$^{-1}$)\\ 
    	\midrule
         $5.581^{+0.020}_{-0.022}$  &   $258^{+165}_{-171}$ &   $2.67^{+0.65}_{-0.57}$   &  \multirow{4}{*}{$6.97^{+0.48}_{-0.48}$}    & \multirow{4}{*}{$0.133^{+0.021}_{-0.020}$}  \\
         $5.787^{+0.024}_{-0.032}$  &   $267^{+158}_{-162}$  &   $1.01^{+0.68}_{-0.71}$   &  &   \\
         $5.901^{+0.045}_{-0.045}$  &   $263^{+158}_{-162}$  &   $0.08^{+0.34}_{-0.06}$   &      &   \\
         $6.042^{+0.045}_{-0.043}$   &   $250^{+169}_{-166}$  &   $0.06^{+0.20}_{-0.04}$   &      &   \\
    \midrule
    \multicolumn{5}{c}{$N_T$ (Total): $3.83^{+1.00}_{-0.91}\times10^{10}$\,cm$^{-2}$} \\
    \bottomrule
    \end{tabular}
    \label{tab:MCMC_fits}
\end{table}

The resulting parameters from the MCMC inference to \ce{$E$-1-C4H5CN} emission in the DR4 observational data are shown in Table~\ref{tab:MCMC_fits}; the corner plot, which provides a more holistic, visual depiction of the explored parameter space, is shown in Fig.~\ref{fig:corner4}. In some cases, especially for less-abundant species where there is not a clear detection in one or more of the velocity components, we find that a three-component model has performed better \citep{McGuire:2020bb}. The majority of the \ce{$E$-1-C4H5CN} column density is detected in two velocity components so we also ran MCMC inference assuming only two velocity components, the results of which can be seen in Fig.~\ref{fig:corner2}, along with the resulting stacked spectrum and matched-filter response (Fig. ~\ref{fig:stack2}). We find both models result in similar total column densities. 

The total column density derived from the DR4 data, assuming a four velocity component model is $3.83^{+1.00}_{-0.91}\times 10^{10}$ cm$^{-2}$. We find a total column density of $3.94^{+0.87}_{-0.79}\times 10^{10}$ cm$^{-2}$ when using two velocity components in our MCMC analysis. The radial velocities, modeled source sizes, excitation temperature, and line width are consistent with similar molecules (that are optically thin) detected in TMC-1. Using these posterior parameters, we generated a spectral stack and performed a matched filtering analysis as described above.  The results are shown in Fig.~\ref{fig:stack}.  The evidence for a detection of \ce{$E$-1-C4H5CN} in our data, as observed in the peak impulse response of the matched filter is 5.1$\sigma$.

\subsection{Astrochemical Modeling}
\label{sec:model}
To study the formation of \ce{1-C4H5CN} and related species, we adapted the three-phase chemical network model \texttt{NAUTILUS} v1.1 code \citep{Ruaud:2016}. Originally based on the \textsc{KIDA} network, \edit1{previous GOTHAM analyses expanded the reaction network to include numerous aromatic and carbon-chain species detected using GOTHAM data.} In all these works, the model \edit1{(hereafter referred to as GOTHAM DR1 Model)} was able to reproduce the observed abundances of the new carbon-chain molecules with reasonable accuracy (typically within a factor of $<$2.5) \citep{Xue:2020:L9, McGuire:2020bb, Loomis:2021aa, Shingledecker:2021}. The abundances of cyclic molecules, however, have been systematically under-produced in each iteration of the model \citep{McCarthy:2021aa, Burkhardt:2020aa, McGuire2021}. 

For this work, we introduced two new species into the chemical network: \ce{1-C4H5CN} and \ce{c-C5H5N}. In our models, we do not distinguish between isomers of \ce{1-C4H5CN}, thus \ce{1-C4H5CN} refers to the total abundance of the acyclic 1-cyano-butadiene isomers. The production and destruction routes involving these molecules are summarized in \edit1{Appendix~\ref{apx:rxn}} with the corresponding rate coefficients at 10\,K. Low-temperature formation routes to \ce{1-C4H5CN} and \ce{c-C5H5N} are not well known. One potential formation route is via dissociative recombination (DR) reactions between N-bearing hydrocarbon ions and electrons, which have been shown to dominate the formation of many carbon-chain molecules in cold environments. However, the formation of the precursor ions, such as \ce{C5H6N+}, is not well understood. Alternatively, laboratory studies have shown that CN radicals react efficiently with hydrocarbons to form the CN-substituted nitriles under low-temperature conditions \citep{Sims1993,Cooke:2020we}, as has been proposed for many CN-substituted hydrocarbons discovered in TMC-1. In this study, we assume the barrierless neutral-neutral reaction of \ce{CH2CHCHCH2} and the \ce{CN} radical to be the only formation path for the two new species, i.e:
\begin{align*}
        \ce{CH2CHCHCH2  +   CN  &-> 1-C4H5CN + H}     &99\%\\
                                &\ce{-> c-C5H5N + H}  &1\%.\\
\end{align*}
\citet{Morales:2011gl} measured the total rate coefficient from room temperature down to 23 K, and gave an expression for its temperature dependence as a modified Arrhenius expression, 
\begin{equation}
    k(T) = (4.8 \pm 0.1) \times 10^{-10} e^{\frac{-77 \pm 10}{8.31 \times T}} \mathrm{cm^3\ s^{-1}}
    \label{eqn:k}
\end{equation}
where $T$ is the gas temperature, in Kelvin. Experimental and computational studies further confirmed that the dominant product is \ce{1-C4H5CN} with a possible minor amount ($<1\%$) of \ce{c-C5H5N} and an absence of 2-cyano-1,3-butadiene \citep{2014JPCA..118.7715S, 2015PCCP...1732000P}. As such, we assumed the branching fraction leading to \ce{1-C4H5CN} to be 99\%. Other potential routes to \ce{c-C5H5N} are discussed in section \ref{sec:pyridine}.

The destruction of \ce{1-C4H5CN} and \ce{c-C5H5N} were assumed to be analogous to that of cyanopolyynes under the TMC-1 conditions \citep{woon_quantum_2009}. The most prevalent destruction pathways in the model are reactions with abundant \edit1{cations (Appendix~\ref{apx:rxn})}. The reaction rate coefficients of the related ion-molecule reactions were estimated with capture rate theory \citep{woon_quantum_2009} and Equation (1) from \citet{Xue:2020:L9}. The dipole moment ($\mu$) and average dipole polarizability ($\alpha$) are $\mu=4.814\ \mathrm{D}$ and $\alpha = 11.178\ \text{\AA}^3$ for \ce{1-C4H5CN} \citep{Zdanovskaia2021}, and $\mu = 2.190\ \mathrm{D}$ \citep{nelson1967selected} and $\alpha = 9.493\ \text{\AA}^3$ \citep{gray1984theory} \edit1{for \ce{c-C5H5N}.}

\edit1{For this study, the physical} conditions of the model are assumed to be consistent with the previous modeling work of TMC\nobreakdash-1 as part of the GOTHAM survey, originally constrained by \citet{hincelin_oxygen_2011}, with a gas and grain temperature of $T_{\text{gas}}$=$T_{\text{grain}}$=10\,K, a gas density of $n_{\rm H}$=2$\times$10$^4$\,cm$^{-3}$, a visual extinction of $A_{\rm v}$=10, and a cosmic ray ionization rate of $\zeta_{\text{CR}}$=1.3$\times$10$^{-17}$\,s$^{-1}$. We adopted the initial elemental abundances described in \citet{hincelin_oxygen_2011} with the exception of atomic oxygen. Through fitting the peak abundances of the cyanopolyyne family to the observed column densities, we previously determined that an \ce{O} element abundance of $1.55 \times 10^{-4}$(/H), corresponding to a C/O ratio of 1.1, best reproduces our observations \citep{Loomis:2021aa}. The resulting molecular abundances, with respect to \edit1{an assumed} $N_\mathrm{T, (\ce{H2})}$ \edit1{=} $1 \times 10^{22} \mathrm{cm^{-2}}$ \citep{Gratier:2016fj}, were converted to column densities and compared with the observed values. \edit1{This $N_\mathrm{T, (\ce{H2})}$ value is consistent with the $\ce{H2}$ column density map of TMC-1 derived from the dust continuum emission observed by \textit{Herschel} \citep{2023MNRAS.519..285S}}.

\begin{figure}
    \centering
    \includegraphics[width=0.45\textwidth]{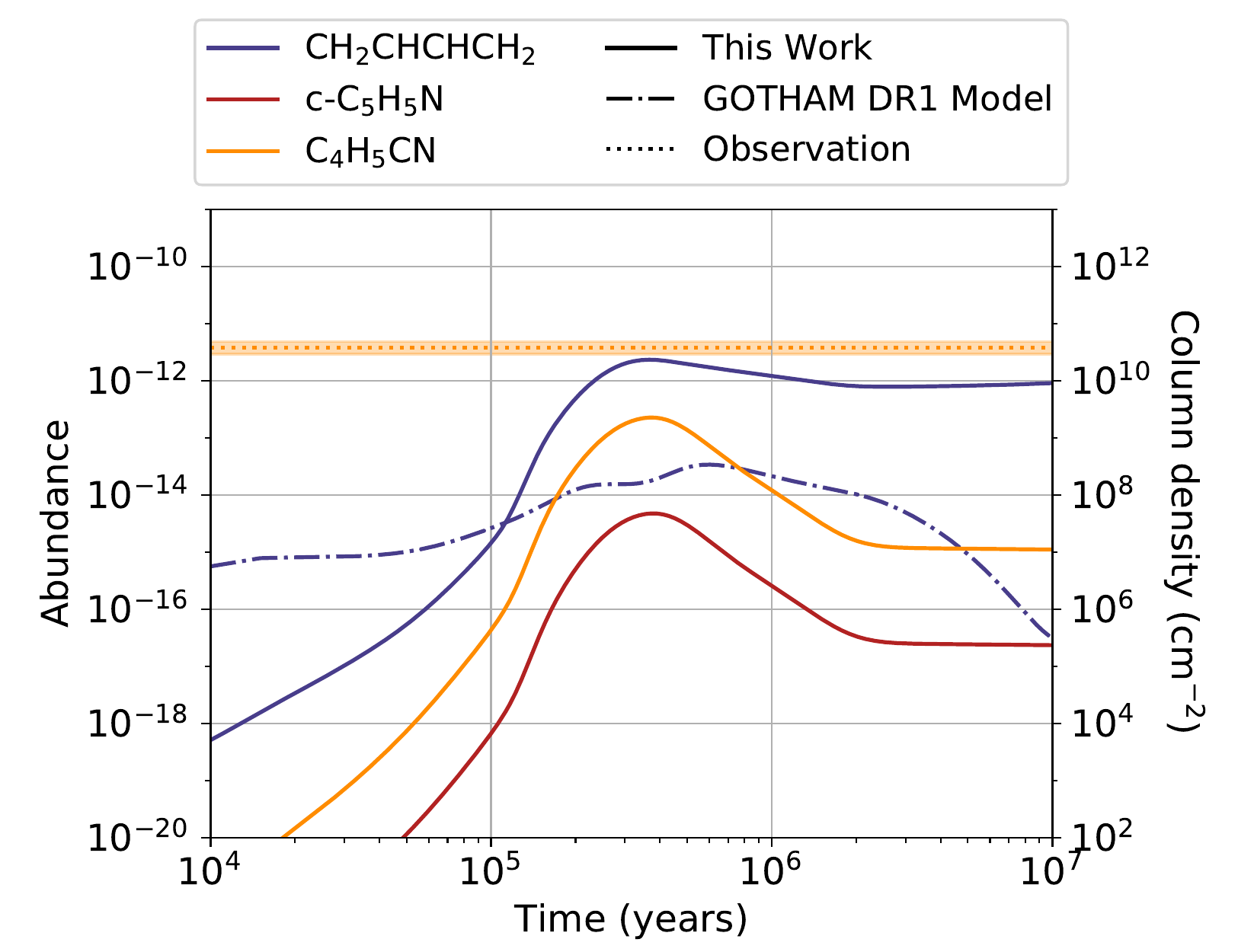}
    \caption{\label{fig:buta} Models of \ce{CH2CHCHCH2} (\edit1{blue}), \ce{1-C4H5CN} (\edit1{orange}) and \ce{c-C5H5N} (\edit1{red}) under the TMC-1 condition. \edit1{Model results obtained with the modified network presented in this work are shown with solid traces, whereas those obtained with GOTHAM DR1 are shown with a dashed-dotted trace.} The observed abundance of \ce{1-C4H5CN} is indicated by the dotted dark purple line along with the uncertainties.} 
\end{figure}

\edit1{In Figure~\ref{fig:buta}, we present} the results of the chemical modeling of \ce{1-C4H5CN}, \ce{c-C5H5N}, and their precursor \ce{CH2CHCHCH2}. The model at an age of ${\sim}3.7\times 10^{5}$ yr has \edit1{a peak abundance of 2.28~$\times 10^{-13}$(/\ce{H2})} for \ce{1-C4H5CN}, $5.9^{+1.4}_{-1.6}\%$ of the observed abundance. In addition, the model predicted \edit1{an abundance of 4.8~$\times 10^{-15}$(/\ce{H2})} for \ce{c-C5H5N} and \edit1{2.34~$\times 10^{-12}$(/\ce{H2})} for \ce{CH2CHCHCH2}. \edit1{Although the current model under-predicts the observed column density of \ce{1-C4H5CN}, the updates we have made here to the reaction network of the related species have shown substantial progress in bringing the models into alignment with observations. In this case, the modeled abundance of \ce{CH2CHCHCH2} (blue solid trace) is increased by about two orders of magnitude compared to the results obtained with the GOTHAM DR1 model (blue dash-dotted trace)}.

\section{Discussion}
\label{sec:dis}
\subsection{Butadiene}

The lack of a permanent dipole moment makes it impossible to detect \ce{CH2CHCHCH2} through its rotational spectra. In contrast, its cyano derivative, \ce{1-C4H5CN}, potentially provides an indirect constraint on the gas-phase abundance of \ce{CH2CHCHCH2} based on the X-CN:X-H ratio.  While the absolute abundance of \ce{1-C4H5CN} predicted by our model still falls well below the observed value, we have recently shown that our models are remarkably good at reproducing the observed ratio X-CN:X-H ratio of CN-substituted:hydrocarbon species.  This was recently demonstrated by \citet{Sita_2022} through the direct observation of the hydrocarbon PAH indene (\ce{C9H8}) and its CN-substituted counterpart, 2-cyanoindene (\ce{C9H7CN}).  A ratio of \ce{C9H8}/\ce{2-C9H7CN} of 43 was observed, which is fairly consistent with the model’s predicted ratio. Comparing the model-predicted ratios against their observed ratios, they found the model was accurate to within a factor of $\sim$2-5  for a range of X-CN:X-H pairs, \edit2{providing evidence that cyano-PAHs may be considered robust observational proxies for their hydrocarbon counterparts.}

%\edit1{\subsubsection{Formation Mechanisms of \ce{CH2CHCHCH2}}}

\edit1{The dominant gas-phase pathways} leading to the formation of \ce{CH2CHCHCH2} at low temperature include the reaction of the methylidyne radical (\ce{CH}) with propene (\ce{CH3CHCH2}),
\begin{equation}
    \label{eqa:C3H6-CH}
        \ce{CH3CHCH2  +   CH  -> CH2CHCHCH2 + H}
\end{equation}
and n-propyl radicals (\ce{CH2CH2CH3}) reacting with atomic carbon,
\begin{equation}
        \label{eqa:C3H7-C}
        \ce{CH2CH2CH3  +   C  -> CH2CHCHCH2 + H}.
\end{equation}

The total rate constant of the reactions between \ce{CH3CHCH2} and CH was measured to be $3.86--4.58 \times 10^{-10}$~cm$^3$ molecule$^{-1}$ s$^{-1}$ over a temperature range of 70--170\,K \citep{B506096F} and $4.2\pm0.8 \times 10^{-10}$~cm$^3$ molecule$^{-1}$ s$^{-1}$ at 298\,K \citep{2009PCCP...11..655L}. While \citet{2009PCCP...11..655L} reported a branching ratio of $(78\pm10)\%$ for the H atom production channel, another recent experiment reported a branching ratio of $(63\pm13)\%$ for the \ce{CH2CHCHCH2} channel at 298\,K \citep{2013JPCA..117.6450T}. Considering that Trevitt et al. did not constrain the total rate constant, we use a rate constant of $3.3 \times 10^{-10}$~cm$^3$ molecule$^{-1}$ s$^{-1}$ for Reaction (\ref{eqa:C3H6-CH}) in our models to be consistent, following the suggestion of \citet{2009PCCP...11..655L} and \citet{Loison:2017}. Additional measurements of the overall rate coefficient for this reaction down to $\sim$10\,K would be valuable. 

Unlike in the gas phase, it seems unlikely that Reaction~(\ref{eqa:C3H6-CH}) would take place on the grain surface. Based on theoretical calculations, the potential energy surfaces for Reaction~(\ref{eqa:C3H6-CH}) proposed by \citet{2009PCCP...11..655L} depict that the \ce{CH2CHCHCH2} + H and allene (\ce{H2CCCH2}) + methyl radical (\ce{CH3}) channels are the most energetically stabilized. However, the calculations presented in \citet{2009PCCP...11..655L} do not take any surface effect into account. Although the surface effect might not affect the reaction barrier heights so much, we expect it impacts the way the reactants sit on the surface and therefore the possible orientations for reactivity. Furthermore, while gaseous intermediates can cross a potential energy surface if it is overall exothermic, any intermediate in the solid state is likely to dissipate its energy fairly quickly. In this way, the \ce{C4H7} intermediates can stabilize and an isomer of the \ce{C4H7} complex becomes the product of this reaction on grains. This suggests that Reaction~(\ref{eqa:C3H6-CH}) can efficiently produce \ce{CH2CHCHCH2} in the gas phase but may not do so on the grain surface, although there might be other surface pathways.

Carbon atoms have been proposed to react with hydrocarbons without any entrance barriers in interstellar environments \citep[e.g.][]{1997ApJ...477..982K, 2013Icar..222..254C}. Based on capture rate theory and the isomeric Reaction~(\ref{eqa:C3H6-CH}), \citet{Loison:2017} suggested Reaction~(\ref{eqa:C3H7-C}) with a rate constant of $1.6 \times 10^{-10}$~cm$^3$ molecule$^{-1}$ s$^{-1}$. Along with Reaction~(\ref{eqa:C3H7-C}), we also incorporated the gas-phase and grain reactions of the relevant 3-carbon hydrocarbons, namely propyne (\ce{CH3CCH}), propenyl radical (\ce{CHCHCH3}), propene (\ce{CH3CHCH2}), n-propyl radical (\ce{CH3CH2CH2}), and propane (\ce{CH3CH2CH3}), listed in \citet{2016MolAs...3....1H} and \citet{Loison:2017}. In particular, dust grain chemistry plays an important role in the formation of precursors to \ce{CH2CHCHCH2}, i.e., the hydrogenation of 3-carbon hydrocarbons \citep{2016MolAs...3....1H}.

These 3-carbon hydrocarbons hydrogenate so efficiently on grain surfaces that their gas-phase abundance is significantly affected by the desorption mechanism and related binding energies. The binding energies ($E_b$) of closed-shell molecules can be probed using temperature-programmed desorption (TPD) experiments. A value of $E_b$ = 2500~K for \ce{CH3CCH} from itself (i.e. in the multilayer regime) was extracted from kinetic modelling of experiments involving reactions between O atoms and \ce{CH3CCH} \citep{2014FaDi..168..167K}. Using the same method, \ce{CH3CHCH2} was found to have a $E_b$ of similar magnitude \citep[2580~K,][]{2011ApJ...741..121W}. Recently, \citet{2019ApJ...875...73B} reported a mean $E_b = 4400$\,K for \ce{CH3CCH}, 3800\,K for \ce{CH3CHCH2}, and 3500\,K for \ce{CH3CH2CH3} from amorphous solid water (ASW) ice. These values are significantly higher than those that have been computed for $E_b$ on model water substrates, e.g., $E_b$(\ce{CH3CCH}) = 2342\,K from water tetramers, $E_b$(\ce{CH3CCH}) = 3153\,K from water hexamers, and $E_b$(\ce{CH3CH2CH3}) = 1456\,K from water monomers \citep{2018ApJS..237....9D, 2022MNRAS.515.3524S}. We chose to use the values reported by \citet{2019ApJ...875...73B} in our models since the authors derived the binding energies directly from TPD analysis from ASW ice, which is likely representative of the ice surfaces present in TMC-1. We find that the higher binding energies measured by \citet{2019ApJ...875...73B} decelerated the release of \ce{CH2CHCHCH2} and its precursors into the gas phase, compared to using the lower binding energies recommended from theoretical studies.  In the case of radical species, we used the computed binding energies listed in \citet{2017MolAs...6...22W}, which are 3100~K for both \ce{CHCHCH3} and \ce{CH3CH2CH2}. As there is no information on the binding energies of \ce{CH2CHCHCH2}, \ce{1-C4H5CN}, and \ce{c-C5H5N}, we used the additive law to estimate those values \citep{2007ApJ...668..294C}, i.e., $E_b(\ce{butadiene}) = E_b(\ce{C2H3}) + E_b(\ce{C2H3}) = 5600\ \mathrm{K}$.

\subsection{Isomer-specific chemistry}

It is important to reiterate that our model does not distinguish between isomers of \ce{1-C4H5CN} in our model. \citet{Morales:2011gl} measured the overall rate coefficient for the reaction of \ce{CN} + \ce{CH2CHCHCH2} by following the time dependent signal from the CN radical via laser-induced fluorescence.  While no information is obtained about the nature of the reaction products using this technique, the authors also conducted crossed molecular beam experiments and measured a branching ratio of $\sim$99\% to the \ce{1-C4H5CN} product channel; however, these experiments do not distinguish between the \textit{E-} and \textit{Z-} isomers. Because our observations only measure the abundance of the E-isomer, and our model represents the sum of the two isomers, care should be taken before direct comparisons are made.  Further observational constraints on the \textit{Z-}isomer, and laboratory constraints on the branching ratio would be highly valuable. 

As for butadiene, a source of uncertainty in its production and reactivity is the role of its minor conformational isomer \textit{gauche}-butadiene, which lies about 2.9~kcal/mol higher in energy than the major \textit{trans} conformer \citep{Saltiel2001}. Theoretical studies have focused almost exclusively on \textit{trans}-butadiene and most experimental studies have been unable to distinguish between the \textit{trans} and \textit{gauche} conformers. In TMC-1, the \textit{trans-gauche} ratio is likely kinetically controlled because of the 3~kcal/mol barrier for the isomerization of \textit{gauche} to \textit{trans}, and therefore the \textit{gauche} conformer might represent a non-negligible portion of the total butadiene abundance. Although the total neutral-neutral reactivity of \textit{gauche}-butadiene with radicals like \ce{C2H} and \ce{CH} can reasonably be speculated to be similar to that of \textit{trans}, the product branching ratios, particularly those to cyclic species, might differ considerably given that the \textit{gauche} geometry is already close to a planar, cyclic carbon framework \citep{Baraban2018}.

The unknown kinetics of \textit{gauche}-butadiene might also contribute to the discrepancies between the flow reactor \citep{Lockyear:2015dn} and crossed molecular beam \citep{Jones:2011yc} laboratory measurements of the \ce{CH2CHCHCH2 + C2H} reaction. The former experiments take place in thermal conditions at 298~K under which the \textit{gauche} population is about 1\%. The \textit{gauche} population in the molecular beam experiments is unknown. Other work has shown that the butadiene conformational temperature can be highly out of equilibrium with the translational and rotational temperatures in a supersonic expansion \citep{Baraban2018}. Although the rest frequencies of \textit{gauche}-butadiene are known, its small dipole moment (ca. 0.1~D) makes its astronomical detection difficult. Based on the current DR4 data set, we can only place a conservative upper bound of $1.28\times10^{11}$ cm$^{-2}$. As with \textit{trans}-butadiene, the cyano derivative presents a likelier detection candidate. The rest frequencies of the \textit{gauche} conformers of \ce{1-C4H5CN} are presently unknown. Follow-up laboratory measurements are thus necessary to enable a search in TMC-1. Understanding what role, if any, \textit{gauche}-butadiene plays in the formation of cyclic species will also require new theoretical investigations and isomer-specific experimental reaction studies.

\subsection{Pyridine}
\label{sec:pyridine}
The detection of \ce{1-C4H5CN} is particularly significant due to its chemical relationship to \ce{c-C5H5N}. \ce{c-C5H5N} is an aromatic heterocyclic molecule of substantial astrochemical and astrobiological interest that has thus far eluded detection in the ISM \citep{Batchelor1973,Charnley2005}, including in our own search \citep{Barnum2022}. \ce{c-C5H5N} is substantially less polar than \ce{1-C4H5CN} ($\mu$ = 2.2 D and $\mu$ = 4.6 D, respectively), which makes it more difficult to detect. The astrobiological relevance of \ce{c-C5H5N} stems from its chemical similarity to pyrimidine (1,3-diazabenzene; \ce{c-C4H4N2}), which itself is a precursor to the DNA nucleobases cytosine, thymine and uracil. In addition, key biomolecules nicotinic acid (vitamin B3) and nicotinamide are composed of functionalized \ce{c-C5H5N} rings. Nicotinic acid has been detected in the Murchison meteorite \citep{Pizzarello2004,Pizzarello2005}, which implies the possibility of its origin in interstellar space. A possible route to nicotinic acid from \ce{c-C5H5N} in interstellar ices has been suggested \citep{McMurtry2016}. 

It is currently unknown whether the cyano-butadienes are chemically linked to \ce{c-C5H5N}, as precursors or decomposition products. \ce{c-C5H5N} is the lowest energy isomer (B3LYP/6-311+G(2d,p); \citet{Zdanovskaia2021}), $\sim$23.2 kcal/mol lower in energy than \ce{1-C4H5CN}, the lowest energy acyclic isomer.  \ce{1-C4H5CN} is itself 0.3 and 16.3 kcal/mol lower in energy than Z-1-cyano-1,3-butadiene and 4-cyano-1,2-butadiene, respectively.  Identification of \ce{c-C5H5N} in interstellar clouds would be valuable to our understanding of the chemical link between prebiotic molecules in interstellar clouds and those identified in meteorites, as well as our understanding of the formation of polycyclic aromatic nitrogen heterocycles (PANHs). The PANHs 1,4-dihydro(iso)quinoline and (iso)quinoline can be synthesized through reaction of pyridyl radicals with \ce{CH2CHCHCH2}, and thus observational constraints of both of these precursors (or their proxies) are sought. 

It remains unclear whether \ce{c-C5H5N} can form in cold molecular clouds like TMC-1. The reaction of CN with \ce{CH2CHCHCH2} has been proposed; however, crossed molecular beam dynamic studies identified the \ce{1-C4H5CN} isomer as the dominant reaction product, with possible minor fractions of the aromatic \ce{c-C5H5N} isomer. Other mechanisms that have been suggested include the ring expansion of pyrrole (\ce{C4H4NH}) by CH \citep{Soorkia2010}, and the reaction of the cyanovinyl radical (\ce{C3H2N}) with vinyl cyanide (\ce{CH2CHCN}; \citealt{Parker2015}). Additional experiments under low-temperature kinetic conditions would be valuable to elucidate whether the dominant product is anticipated to be the cyclic or acyclic isomer in TMC-1. In addition, branching ratios to the \textit{E-} and {Z-}isomers would help to further constrain our models and inform future searches. Experimental setups are currently being developed, such as the Chirped-pulse in Uniform Flow (CPUF) technique \citep{Oldham2014,Abeysekera:2015ge,Hays2020} and CRESU-SOL \citep{Durif2021}, that may be able to shed further light on this reaction in the coming years. 

\section{Conclusions}
\label{sec:con}

We report the detection of \textit{s-trans-E}-1-cyano-1,3-butadiene (\ce{1-C4H5CN}), an acyclic isomer of pyridine (\ce{c-C5H5N}), using the forth data reduction of the GOTHAM deep spectral survey toward TMC-1 with the 100-m Green Bank Telescope. We performed velocity stacking and matched filter analyses using Markov chain Monte Carlo simulations, and find evidence for the presence of this molecule at the 5.1$\sigma$ level. We derive a total column density of $3.8^{+1.0}_{-0.9}\times 10^{10}$ cm$^{-2}$, which is predominantly found toward two velocity components. We use this molecule as a proxy for the apolar hydrocarbon 1,3-butadiene (\ce{CH2CHCHCH2}) and using the three-phase astrochemical model \texttt{NAUTILUS}, we determine a predicted peak column density for \ce{CH2CHCHCH2} of $2.34 \times 10^{10} \mathrm{cm}^{-2}$.  Using the \ce{1-C4H5CN}:\ce{CH2CHCHCH2} ratio derived in our model and the observed column density of \ce{1-C4H5CN}, we further constrain the total column density of 1,3-butadiene in TMC-1 to $3.94 \times 10^{11}\ \mathrm{cm}^{-2}$.  We discuss implications for bottom-up formation routes to aromatic and polycyclic aromatic molecules. 

 %$6.8 \times 10^{11} \mathrm{cm}^{-2}$.
\section{Data access \& code}

Data used for the MCMC analysis can be found in the DataVerse entry \citep{DVN/K9HRCK_2020}. The code used to perform the analysis is part of the \texttt{molsim} open-source package; an archival version of the code can be accessed at \cite{lee_molsim_2020}.

\facilities{GBT}


\software{
    \texttt{NAUTILUS} v1.1 \citep{Ruaud:2016},
    \texttt{Molsim} \citep{lee_molsim_2020},
    \texttt{Emcee} \citep{foreman-mackey_emcee_2013},
    \texttt{ArViz} \citep{kumar_arviz_2019}
          }

\acknowledgments

I.R.C. acknowledges funding from the University of British Columbia, NSERC and the Canada Foundation for Innovation. C.X. thanks V. Wakelam for use of the \texttt{NAUTILUS} v1.1 code. B.A.M. and C.X. gratefully acknowledge the support of NSF grant AST-2205126. The National Radio Astronomy Observatory is a facility of the National Science Foundation operated under cooperative agreement by Associated Universities, Inc.  The Green Bank Observatory is a facility of the National Science Foundation operated under cooperative agreement by Associated Universities, Inc.

%TC:ignore

\bibliography{bibliography}
\bibliographystyle{aasjournal}

\newpage

\appendix

\renewcommand\thefigure{\thesection\arabic{figure}}   
\renewcommand\thetable{\thesection\arabic{table}}    

\setcounter{figure}{0}    
\setcounter{table}{0} 

\section{MCMC Analysis Results}

\begin{figure}
    \centering
    \includegraphics[width=\textwidth]{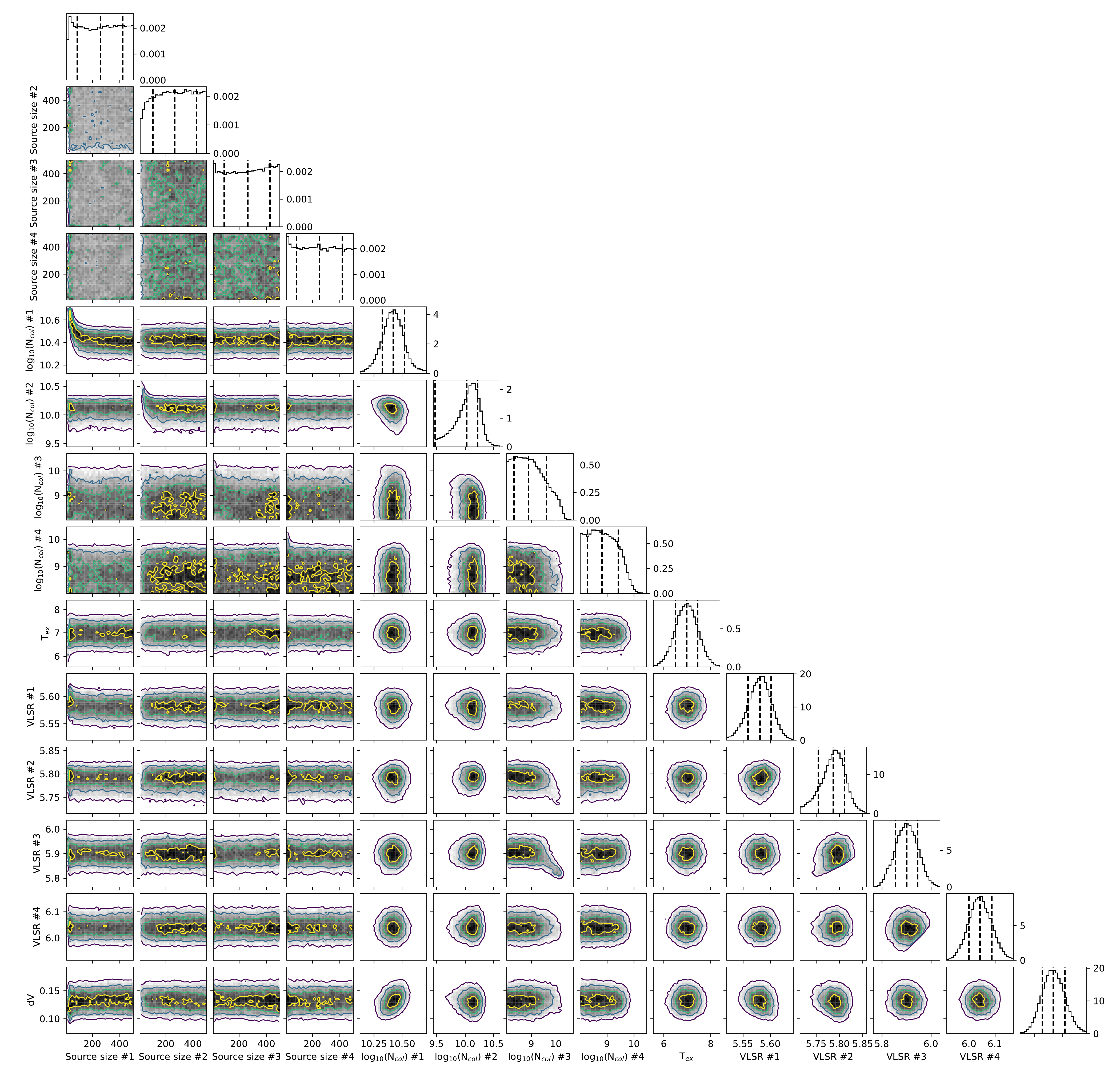}
    \caption{Corner plot for cyano-butadiene showing parameter covariances and marginalized posterior distributions for the MCMC fit. 16$^{th}$, 50$^{th}$, and 84$^{th}$ confidence intervals (corresponding to $\pm$1 sigma for a Gaussian posterior distribution) are shown as vertical lines. }
    \label{fig:corner4}
\end{figure}

\begin{figure}
    \centering
    \includegraphics[width=\textwidth]{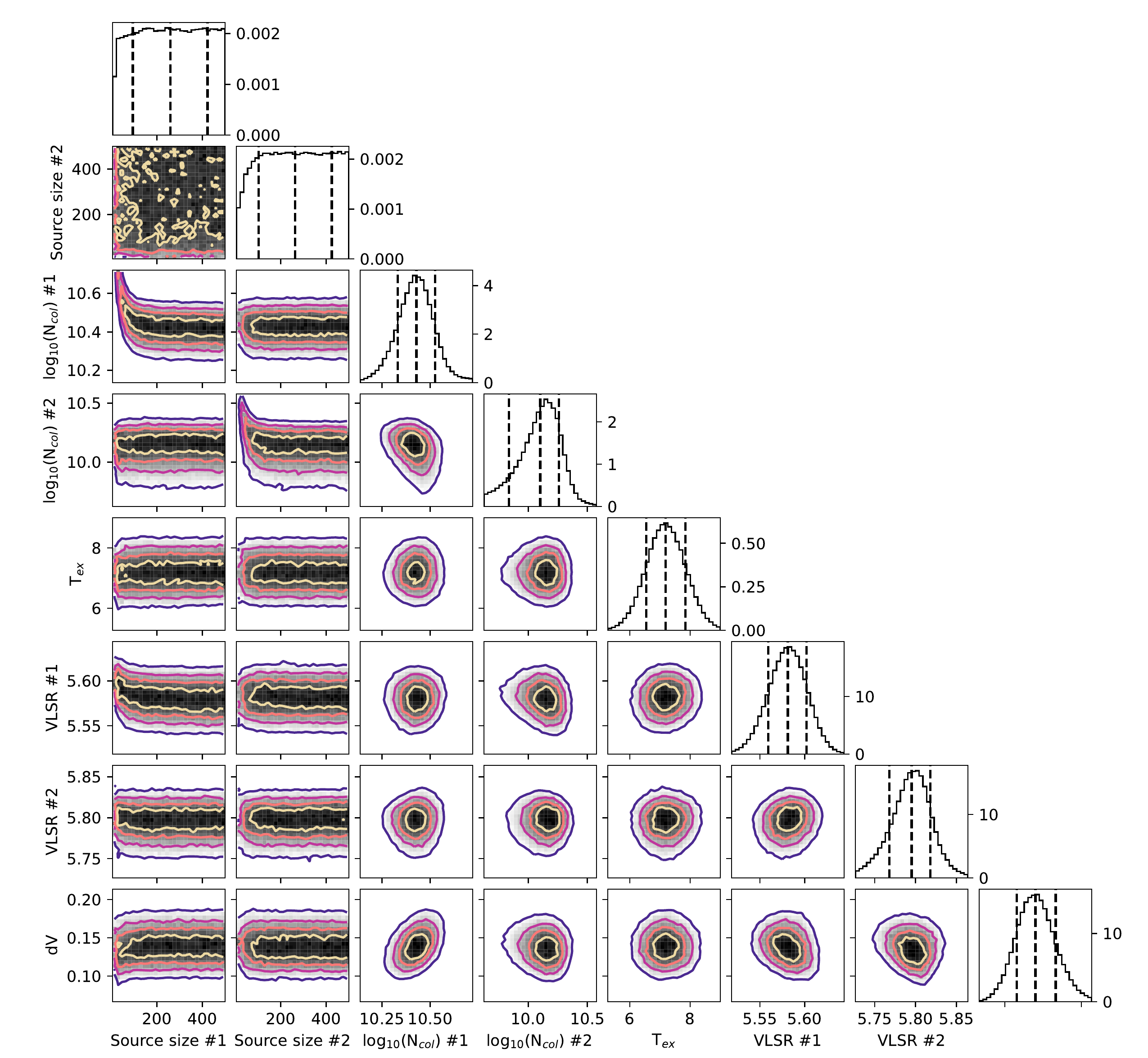}
    \caption{Corner plot for cyano-butadiene showing parameter covariances and marginalized posterior distributions for the MCMC fit. 16$^{th}$, 50$^{th}$, and 84$^{th}$ confidence intervals (corresponding to $\pm$1 sigma for a Gaussian posterior distribution) are shown as vertical lines. }
    \label{fig:corner2}
\end{figure}

\begin{figure*}[bt]
    \centering
    \includegraphics[width=0.45\textwidth]{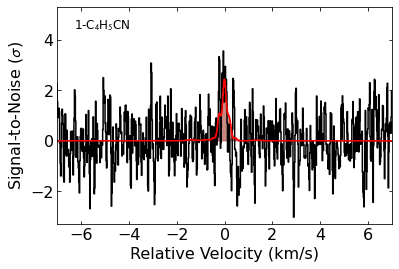}
    \includegraphics[width=0.45\textwidth]{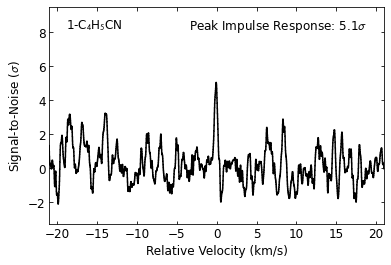}
    \caption{Velocity stacked and matched filter spectra of CN-butadiene, using the 2-component MCMC analysis. The intensity scales are the signal-to-noise ratios (SNR) of the response functions when centered at a given velocity. The ``zero'' velocity corresponds to the channel with the highest intensity to account for blended spectroscopic transitions and variations in velocity component source sizes. (\emph{Left}) The stacked spectra from the GOTHAM DR4 data are displayed in black, overlaid with the expected line profile in red from our MCMC fit to the data.  The signal-to-noise ratio is on a per-channel basis. (\emph{Right}) Matched filter response obtained from cross-correlating the simulated and observed velocity stacks in the left panel; value annotated corresponds to the peak impulse response of the matched filter.}
    \label{fig:stack2}
\end{figure*}

The corner plots resulting from the analysis of cyano-butadiene are shown in Fig.~\ref{fig:corner4} and Fig.~\ref{fig:corner2}, respectively. The velocity-stacked and matched-filter spectra using the posteriors derived from the 2-component model are shown in Figure \ref{fig:stack2}.

\clearpage

\subsection{Spectroscopic Catalogs}

\begin{figure*}
\centering
\includegraphics[width=0.45\textwidth]
{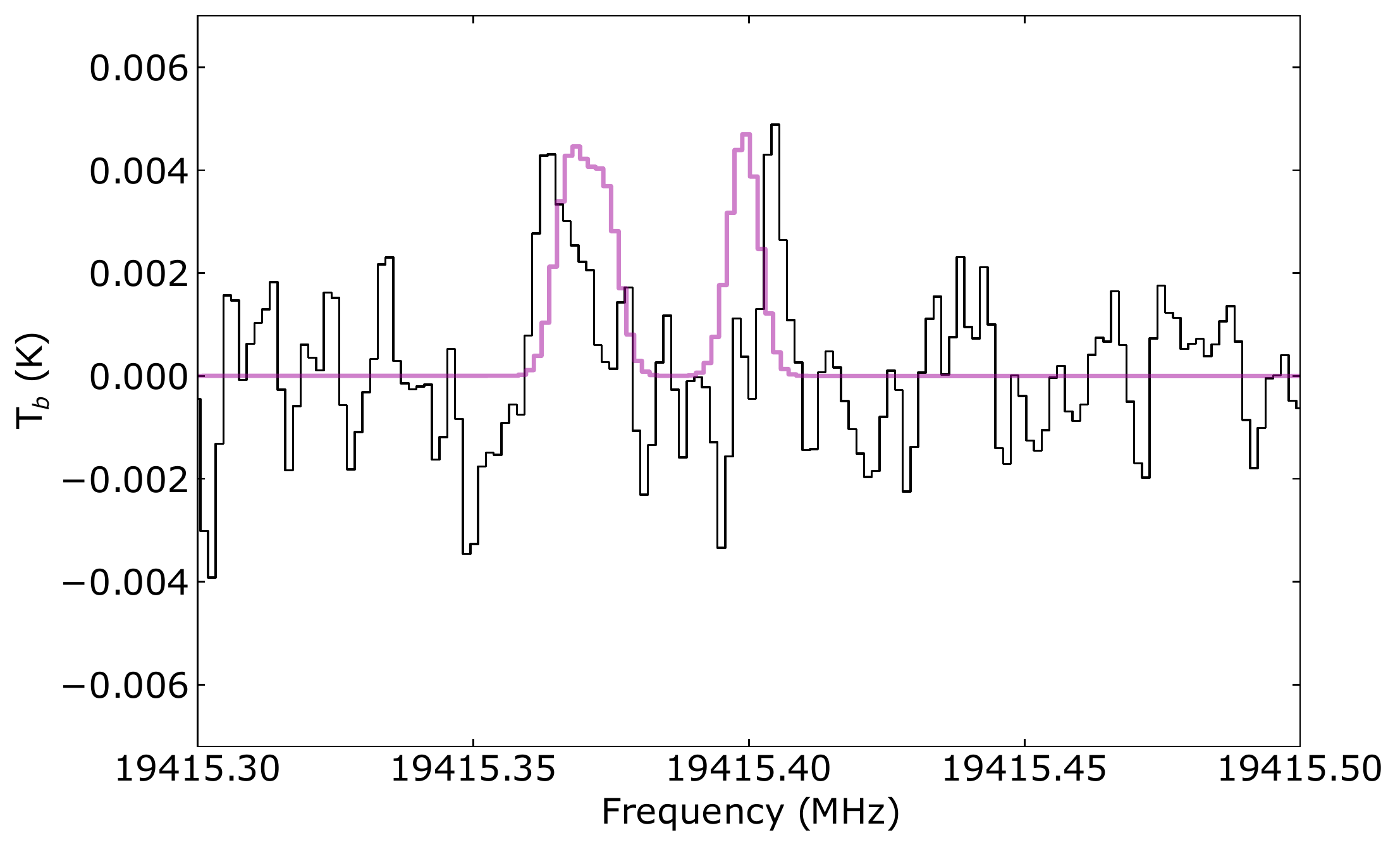}
\includegraphics[width=0.45\textwidth]
{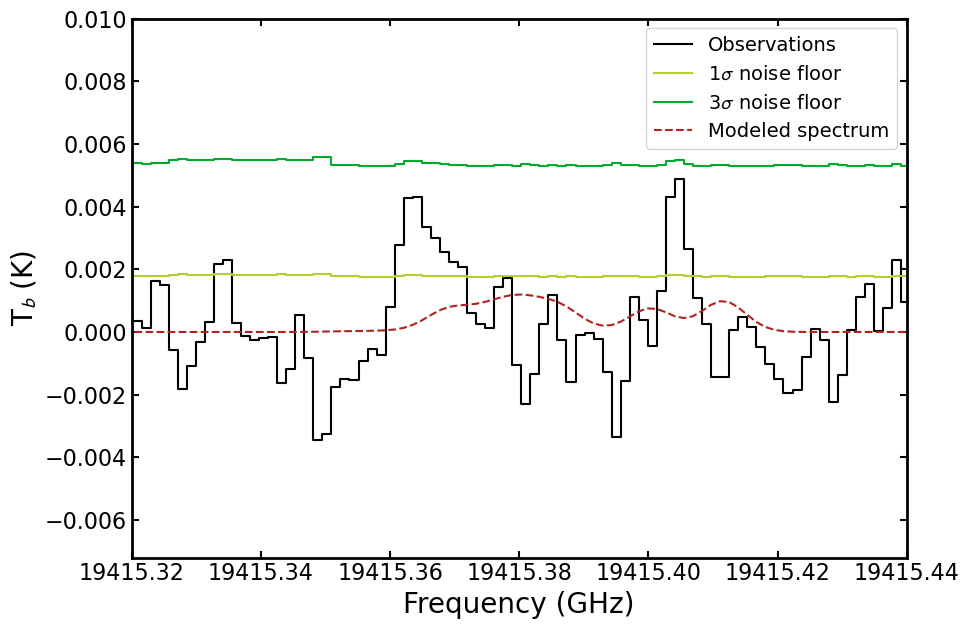}
\caption{\label{fig:weak_lines} (\emph{Left}) Weak interlopers detected in DR4 (black) at $\sim$19415.4 MHz and a simulated single velocity component spectrum (purple) showing the $J = 7_{1,7} \rightarrow 6_{1,6}$ hyperfine transitions reported in \citet{Zdanovskaia2021}. (\emph{Right}) Simulated spectrum (red dashed line) using the outputs of the four-component MCMC analysis. The green and yellow lines indicate the 1$\sigma$ and 3$\sigma$ noise floors, respectively.} 
\end{figure*}

In our initial search for individual lines of the cyano-butadiene isomers, we detected faint transitions at $\sim$19415.4 MHz (Figure \ref{fig:weak_lines}), which are within $\sim$5 kHz of the $J_{K_a,K_c} = 7_{1,7} - 6_{1,6}$ hyperfine transitions of \textit{E}-1-cyano-1,3-butadiene based on the predicted frequencies from laboratory measurements \citep{McCarthy2020,Zdanovskaia2021}. These predictions have an estimated uncertainty of less than 1~kHz, suggesting the astronomical features are interlopers, but because the $7_{1,7} - 6_{1,6}$ transition was not directly observed in the original laboratory studies, this assignment is uncertain.

To resolve this ambiguity, we performed new measurements of these and several other rotational transitions of \textit{E}-1-cyano-1,3-butadiene using the same cavity Fourier-transform microwave spectrometer, discharge expansion source, and conditions as the initial laboratory study \citet{McCarthy2020}, except for switching to a more efficient precursor mix of 1,3-butadiene and acrylonitrile instead of benzene and nitrogen. All three $^{14}$N quadrupole hyperfine components for the previously unreported $7_{1,7} - 6_{1,6}$, $7_{0,7} - 6_{0,6}$,  $7_{1,6} - 6_{1,5}$, and  $5_{1,5} - 4_{1,4}$ rotational transitions were observed. Their rest frequencies are in uniform agreement with the catalog predictions to within the measurement uncertainty (2~kHz). We conclude that the features observed astronomically are indeed interlopers and thus removed the overlapping cyanobutadiene transition from the analysis.

\edit1{\section{Reaction Network \& Uncertainties}}
\label{apx:rxn}

\edit1{The proposed production and destruction routes involving \ce{1-C4H5CN} and \ce{c-C5H5N} are listed in Table~\ref{tab:reactions}. Definitions of $\alpha$, $\beta$, and $\gamma$ can be found on the KIDA online database (\url{http://kida.astrophy.u-bordeaux.fr/help.html}). Formulae of type 3 and 4 are $k(T) = \alpha \left(T/300\right)^\beta e^{-\gamma/T}$ and $k(T) = \alpha \beta \left(0.62 + 0.4767 \gamma \left(300/T\right)^{0.5}\right)$, where $k$ is in $\mathrm{cm^3\,s^{-1}}$ and $T$ is in $\mathrm{K}$, respectively. The corresponding rate coefficients at 10\,K are also listed.}

\edit1{We report the modelled abundances to three digits. The accuracy of the modeled values may be lower with the consideration of uncertainties in the input parameters such as initial conditions and rate coefficients. However, it is infeasible to estimate the accuracy of the modeled values because of, for example, the difficulty in determining the order of accuracy of the nontrivial numerical methods used to solve the rate equations. As such, the presented values were obtained in a realization of the numerical model using the exact values of the input parameters described above. The modeled values in this case are accurate to up to 16 digits, minus the accuracy of the numerical method, which presumably gives results far more accurate than three digits. We present only three digits in order to match the precision of the observed values.}

\begin{deluxetable*}{lccccll}
    \rotate
    % \tablewidth{0pt}
    % \tablewidth{\columnwidth}
    \tablecaption{Summary of the Proposed Reactions For \ce{c-C5H5N} and \ce{1-C4H5CN} \label{tab:reactions}}
    \tablehead{
        \colhead{Reactions} & \colhead{$\alpha$} & \colhead{$\beta$}  & \colhead{$\gamma$} & \colhead{Formula Type} & \colhead{$k (10 \mathrm{K})$} & \colhead{Reference}
    }
    \startdata
        \sidehead{Production Routes:}
        \ce{CN   +  CH2CHCHCH2  ->  H         +  1-C4H5CN             }   &$4.752\times10^{-10}$   &0                      &9.266    &3   &$1.881\times10^{-10}$ &\citep{Morales:2011gl,2015PCCP...1732000P}\\
        \ce{CN   +  CH2CHCHCH2  ->  H         +  c-C5H5N            }   &$4.800\times10^{-12}$   &0                      &9.266    &3   &$1.900\times10^{-12}$ &\citep{Morales:2011gl,2015PCCP...1732000P}\\
        \sidehead{Destruction Routes:}
        \ce{HCO+ + 1-C4H5CN      ->  CO        +  CH3CHCH2   +  C2N+ }   &1.0                     &$1.694\times10^{-9}$   &5.003    &4   &$2.318\times10^{-8}$ &capture rate theory\\
        \ce{H3O+ + 1-C4H5CN      ->  H2O       +  CH3CHCH2   +  C2N+ }   &1.0                     &$1.994\times10^{-9}$   &5.003    &4   &$2.728\times10^{-8}$ &capture rate theory\\
        \ce{H3+  + 1-C4H5CN      ->  H2        +  CH3CHCH2   +  C2N+ }   &1.0                     &$4.589\times10^{-9}$   &5.003    &4   &$6.279\times10^{-8}$ &capture rate theory\\
        \ce{He+  + 1-C4H5CN      ->  He        +  C4H5+      +  CN   }   &1.0                     &$3.998\times10^{-9}$   &5.003    &4   &$5.470\times10^{-8}$ &capture rate theory\\
        \ce{H+   + 1-C4H5CN      ->  H         +  C4H5+      +  CN   }   &0.50                    &$7.851\times10^{-9}$   &5.003    &4   &$5.371\times10^{-8}$ &capture rate theory\\
        \ce{H+   + 1-C4H5CN      ->  H2        +  C5H4N+             }   &0.50                    &$7.851\times10^{-9}$   &5.003    &4   &$5.371\times10^{-8}$ &capture rate theory\\
        \ce{C+   + 1-C4H5CN      ->  C         +  C4H5+      +  CN   }   &0.33                    &$2.417\times10^{-9}$   &5.003    &4   &$1.102\times10^{-8}$ &capture rate theory\\
        \ce{C+   + 1-C4H5CN      ->  CN        +  C5H5+              }   &0.33                    &$2.417\times10^{-9}$   &5.003    &4   &$1.102\times10^{-8}$ &capture rate theory\\
        \ce{C+   + 1-C4H5CN      ->  C2N+      +  CH2CHC2H   +  H    }   &0.33                    &$2.417\times10^{-9}$   &5.003    &4   &$1.102\times10^{-8}$ &capture rate theory\\
        \ce{HCO+ +  c-C5H5N     ->  CO        +  CH3CHCH2   +  C2N+ }   &1.0                     &$1.561\times10^{-9}$   &2.470    &4   &$1.103\times10^{-8}$ &capture rate theory\\
        \ce{H3+  +  c-C5H5N     ->  H2        +  CH3CHCH2   +  C2N+ }   &1.0                     &$4.229\times10^{-9}$   &2.470    &4   &$2.990\times10^{-8}$ &capture rate theory\\
        \ce{H3O+ +  c-C5H5N     ->  H2O       +  CH3CHCH2   +  C2N+ }   &1.0                     &$1.837\times10^{-9}$   &2.470    &4   &$1.299\times10^{-8}$ &capture rate theory\\
        \ce{He+  +  c-C5H5N     ->  He        +  C4H5+      +  CN   }   &1.0                     &$3.685\times10^{-9}$   &2.470    &4   &$2.605\times10^{-8}$ &capture rate theory\\
        \ce{H+   +  c-C5H5N     ->  H         +  C4H5+      +  CN   }   &0.33                    &$7.235\times10^{-9}$   &2.470    &4   &$1.705\times10^{-8}$ &capture rate theory\\
        \ce{H+   +  c-C5H5N     ->  H2        +  C5H4N+             }   &0.33                    &$7.235\times10^{-9}$   &2.470    &4   &$1.705\times10^{-8}$ &capture rate theory\\
        \ce{H+   +  c-C5H5N     ->  CH3CHCH2  +  C2N+               }   &0.33                    &$7.235\times10^{-9}$   &2.470    &4   &$1.705\times10^{-8}$ &capture rate theory\\
        \ce{C+   +  c-C5H5N     ->  C         +  C4H5+      CN      }   &1.0                     &$2.228\times10^{-9}$   &2.470    &4   &$1.575\times10^{-8}$ &capture rate theory\\
    \enddata
\end{deluxetable*}

\end{document}